\documentclass[12pt]{article}
\usepackage{graphicx}

\newsavebox{\hflrar}
\sbox{\hflrar}{\makebox[0pt][l]
{${\scriptstyle \leftharpoonup}$}{${\scriptstyle \rightharpoonup}$}}
\def \hflrpt {\raisebox{1.8ex}{\usebox{\hflrar}} \hspace*{-8pt} \partial^+}

\def \jpsi {J/\psi}
\def \to {\rightarrow}
\def \abst {\vert t \vert}
\def\bfsig{\mbox{\boldmath$\sigma$}}
\def\DT{\mbox{\boldmath$\Delta_T$}}
\def \jpsi {J/\psi}
\def\bfej{\mbox{\boldmath$\varepsilon$}}
\def\FT{T_R \vert_{t\to 0}}

\begin{document}
\begin{flushright}
AS-ITP-2001-018
\end{flushright}
\pagestyle{plain}
\vskip 10mm
\begin{center}
{\Large Soft Gluon Approach for Diffractive Photoproduction of $J/\psi$} \\
\vskip 10mm
J. P. Ma   \\
{\small {\it Institute of Theoretical Physics , Academia
Sinica, Beijing 100080, China }} \\
~~~ \\
Jia-Sheng Xu \\
{\small {\it China Center of Advance Science and Technology
(World Laboratory), Beijing 100080, China }} \\
{\small {\it and Institute of Theoretical Physics , Academia
Sinica, Beijing 100080, China }}
\end{center}

\vskip 0.4 cm


\begin{abstract}

We study diffractive photoproduction of $J/\psi$ by taking the
charm quark as a heavy quark. A description of nonperturbative
effect related to $J/\psi$ can be made by using NRQCD. In the
forward region of the kinematics, the interaction between the
$c\bar c$-pair and the initial hadron is due to exchange of soft
gluons. The effect of the exchange can be studied by using the
expansion in the inverse of the quark mass $m_c$. At the leading
order we find that the nonperturbative effect related to the
initial hadron is represented by a matrix element of field
strength operators, which are separated in the moving direction of
$J/\psi$ in the space-time. The S-matrix element is then obtained
without using perturbative QCD and the results are not based on
any model. Corrections to the results can be systematically added.
Keeping the dominant contribution of the S-matrix element in the large energy
limit we find that the imaginary part of the S-matrix element is related to the
gluon distribution for $x\to 0$ with a reasonable assumption,
the real part can be obtained with another approximation or with
dispersion relation.  Our approach is different
than previous approaches and also our results are different than
those in these approaches. The differences are discussed in
detail. A comparison with experiment is also made and a
qualitative agreement is found. \vskip 5mm \noindent PACS numbers:
12.38.-t, 12.39.Hg, 13.60.-r, 13.60.Le
\par\noindent
Key Words: Soft gluon, HQET, diffractive photoproduction of $J/\psi$,
HERA experiment.
\end{abstract}

\vfill\eject\pagestyle{plain}\setcounter{page}{1}


\par\noindent
{\bf 1. Introduction}
\par\vskip20pt
It is usually believed that the nonperturbative QCD plays an important
role in diffractive processes and one cannot  use perturbative QCD
to describe them. Recently it was pointed out\cite{SJB} that for diffractive
production of a vector meson $V$ like
\begin{equation}
 \gamma^* + h \to h +V
\end{equation}
can be handled with perturbative QCD provided that the virtuality $Q^2$ of
the initial photon is large. This enables us to make testable predictions
for the process and it provides an interesting way to study the nonperturbative
nature of the initial hadron, e.g., the structure function of hadrons and of
nuclei. For $V=J/\psi$ it is also studied
in \cite{MR} with perturbative QCD.
Theoretically it is proved that the S-matrix element
can be factorized\cite{JC}. Neglecting higher orders of $Q^{-2}$ the S-matrix element
consists of the light-cone wave function of $V$, skewed parton distributions
of $h$ and a hard scattering kernel, the hard scattering kernel can be safely
calculated with perturbative QCD and it is free from infrared singularities.
It should be noted that the factorization also holds if one replaces
the vector meson $V$ with a spin-0 meson, or with a photon, the so called
deeply virtual Compton scattering\cite{Ji}.
\par
In this work we study the diffractive photoproduction of $J/\psi$
\begin{equation}
\gamma + h \to h +J/\psi.
\end{equation}
Neglecting a possible $c\bar c$-content of $h$, the process can be imagined as
the following:  The photon splits into a $c\bar c$-pair, after interactions
with the hadron $h$ through gluon exchanges the $c\bar c$-pair is formed
into $J/\psi$. Because the initial photon is real, i.e., $Q^2=0$, the factorization
proved in \cite{JC} does not apply here. If the total energy is sufficiently large,
the exchanged gluons are soft and they can not be handled with perturbative QCD.
But the charm quark can be taken as a heavy quark, for emissions of soft gluons
by heavy quarks the heavy quark effective theory(HQET) can be used\cite{HQET}, a
systematic expansion in the inverse of the charm quark mass $m_c$ can be
employed to study emissions of soft gluons. Taking the charm quark as a heavy quark
it also allows us to use nonrelativistic QCD(NRQCD) to describe nonperturbative
properties of $J/\psi$. As an approximation one can take $J/\psi$ as a bound system
of a $c$- and $\bar c$-quark,
in which the $c$- and $\bar c$-quark has a momentum which is half of the momentum
of $J/\psi$. Corrections to this approximation can be systematically added
in the framework of NRQCD\cite{nrqcd}.
\par
The process studied here has a close similarity to the decay of $J/\psi$
\begin{equation}
 J/\psi \to \gamma^* + \pi+\pi \to e^++e^- +\pi +\pi
\end{equation}
in the kinematic region where the pions are soft. This decay is studied
in \cite{Ma,MaXu}. The exchange of soft gluons
between the $c\bar c$ pair in $J/\psi$ and the pion pair is responsible for the
decay. In \cite{Ma} it is shown that one can use a technique of path integral
for the exchange of soft gluons without invoking perturbative QCD. It is also
shown that at the leading order of $m_c^{-1}$ the decay amplitude derived
with the technique of path integral is the same
as that derived by assuming that the exchange is of two soft gluons in a
special gauge. In this work we will take a suitable gauge and assume
the two-gluon exchange in the gauge to derive the S-matrix element for
the diffractive process. The assumption can be justified by using HQET:
In a suitable gauge the probability for a $c$-quark emitting 1 or 2 gluons is
proportional to $m_c^{-1}$, while the probability for emission of more than 2 gluons
is at order of $m_c^{-n}$ with $n>1$. It is interesting to note our result can be derived
without taking a suitable gauge and the assumption of two gluon exchange.
This
can be done with the technique
of path integral as that derived for the decay in \cite{Ma}, we briefly sketch how to derive our
result in this way and details may be found in \cite{Ma}. Because
the result is derived without the assumption of two gluon exchange in a suitable
gauge, our result actually includes effects of exchange of more than 2 gluons in an
arbitrary gauge. This will be discussed in detail after our result is represented.
The technique of path integral has been used to study
exchanges of soft gluons between two light quarks with large momenta\cite{Na}, where
the exchanges were responsible for diffractive scattering of light hadrons.
\par
Our results for the S-matrix element consists of a NRQCD
matrix element and a distribution amplitude of gluons in the light hadron
$h$. The NRQCD matrix element represents the nonperturbative effect related
to $J/\psi$, the distribution amplitude is defined by a matrix element
of two field strength operators separated in the moving direction of $J/\psi$
in the space-time. It should be emphasized that the obtained
results are not based on any model, corrections to the results
can be systematically added in the framework of QCD.
At the leading order we consider, the produced $J/\psi$ has
the same polarization of the photon, i.e., the produced
$J/\psi$ is transversely polarized.
In the limit of large beam energies, i.e., $s\to\infty$,
the dominant contribution of the amplitude is related to the skewed gluon distribution of $h$.
With a reasonable assumption the forward S-matrix element can be related to the usual gluon
distribution $g_h(x)$ for $x\to 0$. This enables us to predict the forward
differential cross section with available information of $g_h(x)$. With this
result it provides an interesting way in experiment
to access the small $x$-region of $g_h(x)$.
\par
There exist two approaches for the diffractive process in Eq.(1). One is to use
perturbative QCD\cite{SJB,MR}, with several approximations one obtains the S-matrix element
related to the gluon distribution. Another one is based on the perturbative QCD
result for interaction of a small transverse-size dipole of a quark pair, which
is formed into $V$. The interaction of the dipole with the initial hadron
is through two-gluon exchange\cite{dipol}. Both approaches have close similarities,
In these approaches all hadrons including $J/\psi$ should be taken as massless
at the leading order of $Q^{-2}$. Because an expansion in $Q^{-2}$ is used,
one can not take the limit $Q^2\to 0$ to obtain predictions for the photoproduction.
The expansion in $Q^{-2}$ also implies that the S-matrix element is obtained
in the limit of $s\to\infty$ since $Q^2/s$ remains finite.
In \cite{FKS} the approach of the dipole interaction\cite{dipol}
is used, and the effect of the nonzero mass of $J/\psi$ is taken into account. It is found that
one can take $Q^2\to 0$ to have the S-matrix element for the photoproduction,
where the S-matrix element is also related to the gluon distribution.
Similar results are also obtained in \cite{MR}. It is questionable if the
limit $Q^2\to 0$ can be taken because the higher orders of $Q^{-2}$ are
neglected.
Our approach is distinctly different. We start directly from the process in Eq.(2)
and obtain the S-matrix element for a moderate $s$. Then we take the limit
$s\to\infty$, and the forward S-matrix element in the limit is related to the usual
gluon distribution. Our results are also different than those given in \cite{MR,FKS}.
We will discuss the differences in detail.
\par
At first look, one may generalize our results to the case where the initial
photon is virtual with a small $Q^2$. For transversely polarized $J/\psi$
the generalization is straightforward. But, for longitudinally polarized
$J/\psi$ the generalization seems not possible, because the quark mass $m_c$ is
involved in the polarization vector of $J/\psi$, this can spoil
the expansion in $m_c^{-1}$.
The production of longitudinally polarized
$J/\psi$ deserves therefore a further study and we will briefly discuss
the problem.
\par
Our work is organized as the following: In Sec. 2. we introduce
our notations and derive the S-matrix element in the diffractive region
at the leading order of $m_c^{-1}$. The result is derived by taking
the exchange of two soft gluons in a special gauge into account.
We will briefly discuss how to derive it with the technique of path integral.
We will also briefly discuss the problem of the $U(1)$ gauge invariance
and of the production of longitudinally polarized $J/\psi$.
In Sec.3. we derive the forward S-matrix element
in the large energy limit. The S-matrix element is related
to the usual gluon distribution $g_h(x)$ with $x\to 0$. We discuss
the difference between our approach and another approach. In Sec. 4.
we compare our results with experiment. Sec.5 is our summary.


\par\vskip20pt\noindent
{\bf 2. The soft gluon approach}
\par\vskip20pt
We consider the process
\begin{equation}
\gamma(k) + h(p) \to h(p+\Delta) +J/\psi(k-\Delta),
\end{equation}
where the momenta are given in the brackets. The Mandelstam
variables are defined as
\begin{equation}
s=(k+p)^2,\ \ \ t=\Delta^2.
\end{equation}
We study the process in the kinematic region where $\abst$ is at order
of $\Lambda_{QCD}^2$ and each component of $\Delta$ is at order
of $\Lambda_{QCD}$. $h$ is any light hadron whose mass $m$ is at order
of  $\Lambda_{QCD}$. By taking the charm quark as a heavy quark we
have
\begin{equation}
M^2_{J/\psi} \gg \abst, \ \ \ s \gg \abst.
\end{equation}
It should be noted that we do not require that $s\gg M^2_{J/\psi}$,
our result presented in this section can be applied to a wide range of $s$.
But the value of $s$ should be not too small so that $\abst$ can be small enough.
The smallest value of $\abst$ can be approximated by:
\begin{equation}
\abst_{\rm min} = \frac {m^2 M^4_{J/\psi}}{s ( s - M^2_{J/\psi})}
 +{\cal O}(  \frac {m^4}{s})+ {\cal O}(  \frac {m^4}{M^2_{J/\psi}}),
\end{equation}
hence the value of  $s$ should satisfy:
\begin{equation}
s(s-M^2_{J/\psi}) \gg m^2 M^2_{J/\psi}.
\end{equation}
It should be noted that at the threshold we can also have the conditions
in Eq.(6), but the exchanged gluons are not soft, hence, our approach
presented in this work cannot be used for production at the threshold.


\begin{figure}[hbt]
\centering
\includegraphics[width=9cm]{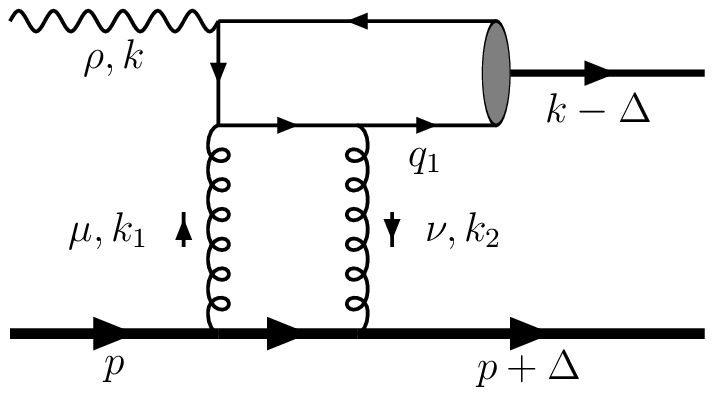}
\caption{One of six Feynman diagrams for elastic photoproduction of
$\jpsi$.}
\label{Feynman-dg1}
\end{figure}

\par
Although our result can be derived exactly by using the technique
of path integral as explained in the introduction, we derive here
our result by assuming two-gluon exchange in a special gauge, because
the derivation in this way is straightforward. We will briefly discuss
how to derive our result without the assumption.
The contributions of two-gluon exchange to the S-matrix element can be represented by
diagrams, one of them is given in Fig.1.  The S-matrix element
with two-gluon exchange can be obtained directly:
\begin{eqnarray}
\langle f \vert S \vert i \rangle &=&\frac{1}{2}
\int d^4x d^4y d^4x_1 d^4x_2 \int \frac{d^4q_1}{(2\pi)^4}
\frac{d^4q_2}{(2\pi)^4}\frac{d^4k_1}{(2\pi)^4}\frac{d^4k_2}{(2\pi)^4}
\nonumber\\
&& \cdot \varepsilon_\rho (k) A_{ij}^{ab,\rho\mu\nu} (k_1,k_2,q_1,q_2)
(2\pi)^4 \delta^4(k+k_1-k_2-q_1-q_2)
\nonumber\\
&& \cdot e^{-iq_1\cdot x -iq_2\cdot y} \langle J/\psi \vert
   \bar c_i(x) c_j(y) \vert 0\rangle
   \cdot e^{ik_1\cdot x_1-ik_2\cdot x_2}
   \langle h(p+\Delta )\vert G^a_{\mu}(x_1) G^b_{\nu}(x_2) \vert h(p)\rangle,
\end{eqnarray}
where $\varepsilon(k)$ is the polarization vector of the photon,
$c(x)$ is the Dirac field of the $c$-quark, the index $i$ and $j$
stands for color- and Dirac indices.
$A_{ij}^{ab,\rho\mu\nu} (k_1,k_2,q_1,q_2)$ is the scattering amplitude
for the process:
\begin{equation}
\gamma(k)+G^*(k_1,a) \to G^*(k_2,b) +c^*(q_1)+\bar c^*(q_2),
\end{equation}
where quarks and gluons are not necessarily on-shell.
The matrix $\langle J/\psi \vert\bar c_i(x) c_j(y) \vert 0\rangle$
represents the nonperturbative effect related to $J/\psi$. It
should be noted that for exchange of arbitrary numbers of gluons
the same matrix appears in the S-matrix element. Because we
take the charm quark as a heavy quark, the $c$- or $\bar c$-quark
in $J/\psi$ carries roughly the half of the momentum of $J/\psi$,
the effect induced by the deviation from the half momentum
is suppressed, the suppression parameter
is the velocity $v_c$ of the $c$- or $\bar c$-quark in $J/\psi$ in
its rest frame. This fact can be realized by boosting the moving frame of $J/\psi$
to its rest frame, in the rest frame one can then uses NRQCD
to perform an expansion in $v_c$\cite{nrqcd}. We will treat
the matrix in the moving frame by using HQET, and then the nonperturbative
effect is represented by matrix elements defined in HQET. These
matrix elements can be related to those defined in NRQCD. Although
the expansion parameter in HQET and in NRQCD is different, but at the
orders we consider here this will not cause problems. In this work
we take nonrelativistic normalization for heavy quark states and
for $J/\psi$ state.
\par
We define the velocity $v$ of $J/\psi$ as
\begin{equation}
  v^\mu = \frac {(k-\Delta)^\mu}{M_{J/\psi}},
\end{equation}
the Dirac field $c(x)$ can be expanded in $m_c^{-1}$ with
fields of HQET:
\begin{equation}
c(x) = e^{-im_c v\cdot x} \left\{ h(x) + \frac {i}{2m_c} \gamma\cdot D_T h(x)
   \right\}
  +e^{+im_c v\cdot x} \left\{ g(x) + \frac {i}{2m_c} \gamma\cdot D_T g(x)
  \right\} +{\cal O}(m_c^{-2}),
\end{equation}
where $D_T^\mu=D^\mu-v\cdot D v^\mu$, $D^\mu$ is
the covariant derivative.
$h(x)$ and $g(x)$ are fields of HQET, $h(x)$ can only annihilate a heavy quark and
$g(x)$ can only create an heavy antiquark. These fields have the property
\begin{equation}
\gamma\cdot v h(x)=h(x), \ \ \ \gamma\cdot v g(x)=-g(x),
\end{equation}
they also depend on the velocity $v$. With these fields the matrix can be written
\begin{equation}
\langle J/\psi \vert\bar c_i(x) c_j(y) \vert 0\rangle
  =e^{im_c v\cdot (x+y)} \langle J/\psi \vert\bar h_i(x) g_j(y) \vert 0\rangle
   +\cdots
\end{equation}
where the $\cdots$ stand for higher orders in $m_c^{-1}$. With the
expansion the $c$- and $\bar c$ quark in $J/\psi$ carries the momentum $m_cv$
plus some residual momentum, the effect of the residual momentum
is represented by the space-time dependence of the matrix element of HQET fields.
Because the effect is small, we can neglect the dependence. We obtain:
\begin{eqnarray}
\langle J/\psi \vert\bar c_i(x) c_j(y) \vert 0\rangle
  &=& -\frac{1}{6}e^{im_c v\cdot (x+y)}\left [\frac {(1-\gamma\cdot v )}{2}
    \gamma\cdot \varepsilon^*(v) \frac{(1+\gamma\cdot v)}{2}\right ]_{ji}
    \cdot \langle J/\psi \vert{\bar h} \gamma\cdot\varepsilon(v)
      g\vert 0\rangle \nonumber\\
      && +\cdots ,
\end{eqnarray}
where the matrix is diagonal in the color-space, the matrix element is of local
fields in HQET, $\varepsilon^* (v)$ is the polarization vector of $J/\psi$.
We will neglect the higher orders represented by $\cdots$ in the above
equations, then the momentum of $J/\psi$ is approximated as $2m_cv$.
The matrix element
$\langle J/\psi \vert{\bar h}(x) \gamma\cdot\varepsilon(v)g\vert 0\rangle$
is related to a NRQCD matrix element defined in the rest frame of
$J/\psi$. The relation reads:
\begin{equation}
\sqrt{v^0} \langle J/\psi \vert{\bar h} \gamma\cdot\varepsilon(v)
      g\vert 0\rangle =-
      \langle J/\psi \vert \psi^\dagger\bfsig\cdot\bfej({\bf v}=0)
      \chi \vert 0\rangle
\end{equation}
where $\psi$ and $\chi$ are NRQCD fields for the $c$- and $\bar c$ quark
respectively. $\sigma_i\ (i=1,2,3)$ is the Pauli matrix. This NRQCD matrix
element can be determined from the leptonic decay of $J/\psi$
\begin{equation}
\Gamma (J/\psi\to e^+e^- ) = \alpha_{em}^2 Q_c^2 \frac {2\pi}{3m_c^2}
  \vert \langle J/\psi \vert \psi^\dagger\bfsig\cdot\bfej({\bf v}=0)
      \chi \vert 0\rangle \vert^2.
\end{equation}
Through examination of contributions of higher orders in Eq.(15) and Eq.(14)
one may find that they are suppressed by $v_c^2$ relatively to the leading order
contribution.
\par
With the result in Eq.(15) the S-matrix element reads:
\begin{eqnarray}
\langle f \vert S \vert i \rangle &=&\frac{1}{2}
\int d^4x_1 d^4x_2 \int \frac{d^4k_1}{(2\pi)^4}\frac{d^4k_2}{(2\pi)^4}
(2\pi)^4 \delta^4(k+k_1-k_2-2m_cv )
\nonumber\\
&& \cdot (-\frac{1}{6\sqrt{v^0}}\langle J/\psi \vert \psi^\dagger\bfsig\cdot\bfej({\bf v}=0)
      \chi \vert 0\rangle)\cdot (-\frac {i}{2}
      eQ_c g_s^2\delta_{ab} R^{\mu\nu} (m_cv,k_1,k_2) )
\nonumber\\
&& \cdot e^{ik_1\cdot x_1-ik_2\cdot x_2}
   \langle h(p+\Delta )\vert G^a_{\mu}(x) G^b_{\nu}(y) \vert h(p)\rangle,
\end{eqnarray}
where
\begin{equation}
-\frac {i}{2}
      eQ_c g_s^2\delta_{ab} R^{\mu\nu} (m_cv,k_1,k_2) = {\rm Tr}\left\{
\varepsilon_\rho (k) A_{ij}^{ab,\rho\mu\nu} (k_1,k_2,m_cv,m_cv)
  \gamma\cdot\varepsilon^*(v) \frac {1+\gamma\cdot v}{2}\right\}.
\end{equation}
Now we take the special gauge
\begin{equation}
v\cdot G(x) =0.
\end{equation}
In the gauge we have:
\begin{equation}
 v_\mu G^{\mu\nu}(x) =v_\mu \frac{\partial}{\partial x_\mu} G^\nu(x),
\end{equation}
where $G^{\mu\nu}(x)$ is the field strength tensor of gluon.
In $R^{\mu\nu} (m_cv,k_1,k_2)$ $k_1$ and $k_2$ are the momenta carried by the two
exchanged gluons, their components are small in comparison with $m_c$,
because the two gluons are soft gluons. Hence $R^{\mu\nu} (m_cv,k_1,k_2)$ can be
expanded in $m_c^{-1}$, a formal expansion in $m_c^{-1}$
leads to:
\begin{eqnarray}
R^{\mu\nu} (m_cv,k_1,k_2) &=& \frac{1}{m_c} \cdot \frac {4 \varepsilon(k)\cdot
    \varepsilon^* (v) }{v\cdot (k_1-k_2) } \cdot g^{\mu\nu}\cdot
    \frac {v\cdot k_1 v\cdot k_2 }{(v\cdot k_1 -i0^+)(v\cdot k_2+i0^+)} +{\cal O}(m_c^{-2})
\nonumber\\
&& + ({\rm terms\ proportional\ to\ }v^\mu\ {\rm or}\ v^\nu),
\end{eqnarray}
where $0^+$ denotes an infinitesimal positive number, which comes from
quark propagators. The leading order is $m_c^{-1}$.
Substituting the result of $R^{\mu\nu} (m_cv,k_1,k_2)$
at the leading order in the S-matrix element in Eq.(18), the integrations of the components
of $k_1$, $k_2$, $x_1$ and $x_2$, which are transverse to $v$, can be easily performed,
and the result for the S-matrix element at the leading order
can be obtained straightforwardly. To present the result we define
\begin{eqnarray}
 F_R(z) &=& g_s^2 \int^{\infty}_{-\infty} \frac {d\tau}{2\pi}
 e^{-i\tau z v\cdot\Delta}
 v_\mu v_\nu \langle h(p+\Delta )\vert G^{a,\mu\rho}(\tau v)
 G^{a,\nu}_{\ \ \rho} (-\tau v) \vert h(p)\rangle,
\nonumber\\
 &\approx &
 g_s^2 \int^{\infty}_{-\infty} \frac {d\tau}{2\pi}
 e^{i\tau z m_c}
 v_\mu v_\nu \langle h(p+\Delta )\vert G^{a,\mu\rho}(\tau v)
 G^{a,\nu}_{\ \ \rho} (-\tau v) \vert h(p)\rangle
\end{eqnarray}
where we have used
\begin{equation}
v\cdot\Delta = -m_c \left ( 1+ \frac{t}{4m_c^2}\right )\approx -m_c.
\end{equation}
The variable $z$ is related to the momentum $k_1$ and $k_2$ by
\begin{equation}
v\cdot k_1= -\frac{1}{2} v\cdot\Delta (1+z),\ \ \  v\cdot k_2 = \frac{1}{2} v\cdot\Delta (1-z).
\end{equation}
The function $F_R(z)$ are zero for $\vert z\vert > z_0$ with
$z_0=(2v\cdot p +v\cdot \Delta)/m_c$
because of the conservation of momentum.
With the function the result for the $S$-matrix element reads:
\begin{eqnarray}
\langle f \vert S \vert i \rangle &=& \frac {2i}{3\sqrt{v^0}} eQ_c (2\pi)^4 \delta^4( k-\Delta -2m_cv)
     \frac {\varepsilon (k)\cdot \varepsilon^*(v)}{m_c}
\nonumber\\
&& \cdot \frac {\langle J/\psi \vert \psi^\dagger\bfsig\cdot\bfej({\bf v}=0)
      \chi \vert 0\rangle}{(v\cdot \Delta)^2} \int
      dz \frac{1}{(1+z-i0^+)(1-z-i0^+)} F_R(z)
\nonumber\\
           &\approx & \frac {2i}{3\sqrt{v^0}} eQ_c (2\pi)^4 \delta^4( k-\Delta -2m_cv)
     \frac {\varepsilon (k)\cdot \varepsilon^*(v)}{m_c^3}
\nonumber\\
&& \cdot \langle J/\psi \vert \psi^\dagger\bfsig\cdot\bfej({\bf v}=0)
      \chi \vert 0\rangle \int
      dz \frac{1}{(1+z-i0^+)(1-z-i0^+)} F_R(z).
\end{eqnarray}
The above results are derived with the assumption of two-gluon
exchange in the gauge $v\cdot G =0$. In this gauge the polarization vectors
of the two exchanged gluons are perpendicular to $v$. To maintain the color-gauge invariance
in other
gauges a gauge link must be supplied between the field strength
operators in Eq.(23), with the gauge link the effect of exchanges of gluons,
whose polarization vectors are proportional to $v$ and whose number is unlimited,
is also included. Our result derived in this way may be unsatisfied, because the
assumption of two-gluon emission sounds that we performed an expansion in
$g_s$ for soft-gluons and we add the gauge link by hand. It is possible that
the $c\bar{c}$ pair emits soft gluons whose polarizations are all proportional to $v$
and this emission is not suppressed by $m_c^{-1}$. This type
of contributions was excluded with the gauge. However our result
can be derived in an arbitrary gauge without the assumption of two-gluon
exchange. The derivation is similar to this for the decay in Eq.(3),
we will briefly describe the derivation here, details can be found in
\cite{Ma}.
\par
Considering the process with exchange of arbitrary number of gluons in
an arbitrary gauge,
the matrix element in l.h.s. of Eq.(15) always appears in
the S-matrix element. With the approximation in Eq.(15), it is
equivalent to consider photoproduction of a $c\bar c$-pair,
where the $c$- and $\bar c$ quark is on-shell and has the same momentum
$m_cv$. Using the standard SLZ reduction formula we related the $S$-matrix
element to Green's functions, which can be calculated with QCD path integral.
Imaging that we perform first the integration over $c$-quark fields, then
the problem is formulated as to solve the wave functions of $c$- and
$\bar c$ quark in the process under a background of gluon fields:
\begin{equation}
\gamma \to c +\bar c.
\end{equation}
The background fields vary slowly with the space-time, reflecting
the fact that the exchanged gluons are soft. The wave functions
can be solved with an expansion in $m_c^{-1}$. At leading order, i.e.,
at order of $m_c^0$, the wave-functions are obtained by multiplying
the wave-functions in the free case with gauge links determined
by $v\cdot G$.
This means
that at the order of $m_c^0$ only those gluons whose polarization is proportional
to $v$, are exchanged. Because of the symmetry of charge conjugation these gauge links
do not lead to any physical effect, i.e., the S-matrix element is zero at $m_c^0$.
Solving the wave-functions at order of $m_c^{-1}$,
one obtains exactly the same results
given in Eq.(26), and the gauge link, which needs to be added by hand
in Eq.(23), is automatically generated. Because the results are derived
without the assumption of two-gluon exchange, they are nonperturbative.
The results indicate that the exchange consists of two gluons, whose
polarizations are transverse to the moving direction of $J/\psi$,
and of any number of gluons, whose polarizations
are proportional to $v$.
\par
Our results show that the produced $J/\psi$ is transversally polarized
at the order we consider, the production of longitudinally
polarized $J/\psi$ is suppressed. As they stand, the results do not
respect to the gauge invariance of electromagnetism. This can be
seen that the S-matrix element in Eq.(26) is not zero
if $\varepsilon(k)$ is replaced by $k$. To study the problem, we note
that the performed expansion in $m_c^{-1}$ is a formal expansion,
the true expansion parameter is $(m_cv\cdot k_1)^{-1}$ or
$(m_cv\cdot k_2)^{-1}$, this can also be realized by inspecting
the HQET lagrangian. To identify the expansion parameter more clearly,
we scale the momenta:
\begin{equation}
 k_1 =\lambda \tilde k_1,\ \ \ k_2=\lambda \tilde k_2,
\end{equation}
where the components of $\tilde k_1$ or $\tilde k_2$ are
${\cal O}(1)$, $\lambda$ is proportional to
$\Lambda_{QCD}$ and is small. We note that the $v$-dependence
in $R^{\mu\nu}$ appears though $v\cdot k_1$ and $v\cdot k_2$,
if we use $k\cdot \varepsilon(k)=(2m_cv+k_2-k_1)\cdot \varepsilon(k)
=0$. We scale these factors as:
\begin{equation}
v\cdot k_1 =m_c \tilde v \cdot \tilde k_1, \ \ \
v\cdot k_2 =m_c \tilde v \cdot \tilde k_2,
\end{equation}
where $\tilde v \cdot \tilde k_1$ and $\tilde v \cdot \tilde k_2$
are ${\cal O}(1)$. Now we can expand $R^{\mu\nu}$ in $\lambda$,
the result reads:
\begin{equation}
R^{\mu\nu}(m_c v,k_1,k_2) = \varepsilon_\rho w_1^{\rho\mu\nu}
      +{\cal O}(\frac{\lambda}{m_c})
 + ({\rm terms\ proportional\ to\ }v^\mu\ {\rm or}\ v^\nu).
\end{equation}
The first term is
identical to the term in Eq.(22), which is at order of $\lambda^0$,
the second term has a length form and is at order of
$\lambda$, this implies that the corrections
to our results in Eq.(26) is at order of ${\cal O}(\frac{\lambda}{m_c})$,
where corrections come not only from the term at order of ${\cal O}(\frac{\lambda}{m_c})$,
but also from emission of more than 2 gluons in the gauge.
With this examination the gauge invariance is violated at order of
${\cal O}(\frac{\lambda}{m_c})$, i.e., at the next-to-leading order which
we neglect, by noting that $k\cdot \varepsilon^*(v) ={\cal O}(\lambda)$.
To restore the gauge invariance, one needs to analyze the contribution
at the next-to-leading order. Retaining only the leading order, the
gauge invariance holds.
\par
Our results show that the produced $J/\psi$ has the same helicity
of the initial photon, i.e., the produced $J/\psi$
is transversally polarized. This is easy to be understood by
noting that the exchanged gluons at the considered order do not
change the helicity of the $c$- or $\bar c$ quark. We can formulate
our results as:
\begin{eqnarray}
\langle f \vert S \vert i \rangle &=& (2\pi)^4 \delta^4( k-\Delta -2m_cv)
         \delta_{\lambda_\gamma\lambda_J} \cdot (-i\frac{2}{3}
         eQ_c ) \langle J/\psi \vert \psi^\dagger\bfsig\cdot\bfej({\bf v}=0)
      \chi \vert 0\rangle
\nonumber\\
     && \cdot \frac{1}{\sqrt{v^0} m_c^3} T_R  \cdot
     \left \{ 1 +{\cal O}(m_c^{-1})+{\cal O}(v_c^2) \right\}
\nonumber\\
   T_R  &=& \int dz
    \frac{1}{(1+z-i0^+)(1-z-i0^+)} F_R(z),
\end{eqnarray}
where $\lambda_\gamma$ and $\lambda_J$ is the helicity of
the photon and of $J/\psi$, respectively. This is our main
result of this section, where the order of errors is also
given. The first error is from neglecting higher orders
in the $m_c^{-1}$ expansion for emission of soft gluons, while
the second is from neglecting the relativistic correction related
to $J/\Psi$.
\par
If the initial photon is virtual and the virtuality $Q^2$ is small,
the exchanged gluons are also soft. In this case our approach can be
used. It is straightforward to generalize our results to the production
of transversally polarized $J/\psi$ with the transversally polarized
photon. But, the generalization may not be done for longitudinally
polarized $J/\psi$ with longitudinally polarized photon. The reason
may be seen from the expansion in $m_c^{-1}$ for $R^{\mu\nu}$ in Eq.(22).
If the photon and $J/\psi$ are transversally polarized, their polarization
vectors have components which are all at order of ${\cal O}(1)$. Then $m_c$
is the only large parameter in $R^{\mu\nu}$ and an expansion in $m_c^{-1}$
can be performed. This fact is used in the expansion in Eq.(30).
If they are longitudinal polarized, their polarization
vectors can have components which are very large in comparison with
${\cal O}(1)$, this may prevents us from an expansion in $m_c^{-1}$
for $R^{\mu\nu}$. It deserves a further study of the production
of longitudinally polarized $J/\psi$.


\par\vskip20pt
\noindent
{\bf 3. The forward S-matrix element in the limit of $s\to\infty$}
\par
\vskip20pt
In this section we discuss the S-matrix element in the limit of
$s\to\infty$. In this limit, the function $F_R(z)$ can be directly
related to the skewed gluon distribution, whose definition can be
found in \cite{JC,Ji}. This can be realized by that in the limit the dominant part
of $v$ is proportional to a light cone vector defined below and the correlator
defined in Eq.(23) can be expanded with operators classified with twist. The
leading order is determined by twist-2 operators. However, the skewed gluon
distribution function is not well known and this prevents us from numerical
predictions. But, as we will see, there is a possibility to relate
the imaginary part of the forward S-matrix element
to the usual gluon distribution under certain approximation
as in the case of the process in Eq.(1)\cite{SJB}.
We will show that the imaginary part can be related to
the gluon distribution with a reasonable assumption and the real part can be estimated
by an approximation, then
we obtain the forward
S-matrix element determined by usual
gluon distribution function $g_h(x)$ of $h$ for $x\to 0$.
Before taking the limit $s\to\infty$, we note that the integral over $z$
in $T_R $ in Eq.(31) can be performed analytically. Because
$F_R(z) =0$ for $\vert z\vert >z_0$, we can extend the integration
of $z$ and exchange the integration of $z$ and that of $\tau$
in $F_R(z)$. Using an contour integration we obtain:
\begin{eqnarray}
T_R  &=& (\pi i)g_s^2 \left\{\int_0^\infty \frac{d\tau}{2\pi}
  e^{-i\tau m_c} +\int_{-\infty}^0 \frac{d\tau}{2\pi}
  e^{+i\tau m_c}\right\}
  v_\mu v_\nu \langle h(p+\Delta )\vert G^{a,\mu\rho}(\tau v)
  G^{a,\nu}_{\ \ \rho} (-\tau v) \vert h(p)\rangle
\nonumber\\
 &=& (2\pi i)g_s^2 \int_0^\infty \frac{d\tau}{2\pi}
 e^{-i\tau m_c}v_\mu v_\nu \langle h(p+\Delta )\vert G^{a,\mu\rho}(\tau v)
  G^{a,\nu}_{\ \ \rho} (-\tau v) \vert h(p)\rangle .
\end{eqnarray}
\par
For convenience we take a coordinate system in which the photon moves
in the $-z$-direction and the hadron $h$ in the $z$-direction. We
introduce a light-cone coordinate system, components of a vector $A$ in this
coordinate system are related to those in the usual coordinate system as
\begin{equation}
 A^\mu =\left( A^+,A^-,{\bf A_T}\right) =
       \left( \frac{A^0+A^3}{\sqrt{2}},\frac{A^0-A^3}{\sqrt{2}},A^1,A^2\right).
\end{equation}
The momenta in the process  can be approximated in the limit
$s\approx 2k^-p^+\to\infty$ as:
\begin{eqnarray}
k^\mu &=& (0,k^-,{\bf 0_T}), \nonumber\\
p^\mu &=& (p^+,\frac{m^2}{2 p^+},{\bf 0_T}) \approx
      (p^+,0,{\bf 0_T}), \nonumber\\
\Delta^\mu &=& \left( \frac{t-M_{J/\psi}^2}{2k^-},
                   \frac {-t-\frac{m^2}{p^+}\Delta^+} {2p^+},
                   \DT\right) \approx \left( -\frac{2m_c^2}{k^-},
                   -\frac{t}{2p^+},\DT\right)
       \nonumber\\
       v^\mu &=& \frac{(k-\Delta)^\mu}{M_{J/\psi}}\approx \left( \frac{m_c}{k^-},
       \frac{k^-}{2m_c},{\bf 0_T} \right)\approx \left( 0,
       \frac{k^-}{2m_c},{\bf 0_T} \right),
\end{eqnarray}
where $v$ is approximated as a light cone vector.
Using these approximated momenta and $M^2_{J/\psi} \approx 4m_c^2 \gg \abst$
$T_R $ can be written:
\begin{equation}
T_R
 \approx (2\pi i)g_s^2 \frac{k^-}{4m_c}\int_0^\infty \frac{d\lambda}{2\pi}
 e^{-i \lambda x_c p^+} \langle h(p+\Delta )\vert G^{a,+\rho}(\frac{\lambda}{2} n)
  G^{a,+}_{\ \ \ \rho} (-\frac{\lambda}{2} n) \vert h(p)\rangle,
\end{equation}
with
\begin{equation}
n^\mu =(0,1,{\bf 0_T}), \ \ \ \  x_c=\frac{2m_c^2}{s},
\end{equation}
where $T_R$ is approximated by the matrix element of the twist-2 operator and
corrections can be parameterized with higher-twist operators and they are suppressed
by large scales like $s$.
We consider the forward case, i.e., $t\to 0$. For $t\to 0$ and $\Delta^\mu\to 0$
the integral in Eq.(35) can be approximated by the replacement:
\begin{eqnarray}
\langle h(p+\Delta )\vert G^{a,+\rho}(\frac{\lambda}{2} n)
  G^{a,+}_{\ \ \ \rho} (-\frac{\lambda}{2} n) \vert h(p)\rangle
  &\approx&
  \langle h(p )\vert G^{a,+\rho}(\frac{\lambda}{2} n)
  G^{a,+}_{\ \ \ \rho} (-\frac{\lambda}{2} n) \vert h(p)\rangle
   =f(\lambda),\nonumber\\
T_R
 &\approx& (2\pi i)g_s^2 \frac{k^-}{4m_c}\int_0^\infty \frac{d\lambda}{2\pi}
 e^{-i \lambda x_c p^+} f(\lambda) ,
\end{eqnarray}
with an assumption which we will discuss later in detail.
In Eq.(37) we introduce the notation $f(\lambda)$ for the forward
matrix element, it has the property $f(\lambda)=f(-\lambda)$.
Again, $f(\lambda)$ also appears in the definition
of the gluon distribution $g_h$ in $h$, in the light cone
gauge we use the definition is\cite{CS}
\begin{eqnarray}
xg_h(x) &=& -\frac{1}{p^+} \int^\infty_{-\infty}
\frac{d\lambda}{2\pi} e^{-i\lambda p^+ x}
 f(\lambda)
\nonumber\\
  &=& -\frac{2}{p^+}\int^\infty_0
\frac{d\lambda}{2\pi} \cos(\lambda p^+ x)
 f(\lambda).
\end{eqnarray}
Therefore the forward S-matrix element can be related to the
usual gluon distribution.
At first look, it seems that there is an ambiguity in relating $T_R$ with
the usual gluon distribution. One
can use the translational covariance to shift the variable
$\lambda$ in Eq. (35),
\begin{eqnarray}
   &&  \langle h(p+\Delta )\vert G^{a,+\rho}(\frac{\lambda}{2} n)
   G^{a,+}_{\ \ \ \rho} (-\frac{\lambda}{2} n) \vert h(p)\rangle  \nonumber \\
   && = e^{-i \lambda x_c p^+}
    \langle h(p+\Delta ) \vert G^{a,+\rho}(0)
       G^{a,+}_{\ \ \ \rho} (- \lambda  n) \vert h(p)\rangle .   \nonumber
\end{eqnarray}
If  one makes the approximation,
\begin{eqnarray}
   \langle h(p+\Delta )\vert G^{a,+\rho}(0)
   G^{a,+}_{\ \ \ \rho} (- \lambda  n) \vert h(p)\rangle
   &\approx&
   \langle h(p )\vert G^{a,+\rho}(0)
   G^{a,+}_{\ \ \ \rho} (- \lambda n) \vert h(p)\rangle
   =f(\lambda),\nonumber
\end{eqnarray}
then one will obtain:
\begin{equation}
 T_R
   \approx  (2\pi i) g_s^2 \frac{k^-}{4m_c}\int_0^\infty \frac{d\lambda}{2\pi}
   e^{-i 2 \lambda x_c p^+} f(\lambda) ,
\end{equation}
i.e., one will get a different result, because $x_c$ in Eq.(37) is replaced
by $2x_c$ in Eq.(39) now. This will result in that the forward S-matrix element
will be related to the gluon distribution $g_h(x)$ at $x=2x_c$ instead of
$g_h(x)$ at $x=x_c$, as we will see below in Eq.(41) and Eq.(51).
Then we will
have two different results of one theory.
In the following,  we will use the approximation Eq. (37) to derive our
results and will demonstrate later that the  approximation leading to
Eq.(39) is not consistent because
some neglected contributions in this approximation  are at the same
order as those kept in the approximation.
\par
To study the relation of $T_R$ to the gluon distribution in detail, we write
$T_R $ as a sum of the integrals:
\begin{equation}
\FT =\pi ig_s^2 \frac{k^-}{4m_c} \left\{
 \int^\infty_{-\infty} \frac{d\lambda}{2\pi}
  +\int^\infty_0 \frac{d\lambda}{2\pi} -\int^0_{-\infty} \frac{d\lambda}{2\pi}
 \right\} e^{-i\lambda p^+x_c} f(\lambda)
\end{equation}
where the first term is simply proportional to the gluon distribution with $x=x_c$,
the second and third
terms can be combined into one integral by using the property $f(\lambda)=f(-\lambda)$:
\begin{equation}
 \FT = -i\pi g_s^2 \frac{s}{8m_c} \left[ xg_h(x)\right] \vert_{x=x_c}
   + \pi  g_s^2 \frac{k^-}{2m_c}
 \int^\infty_0 \frac{d\lambda}{2\pi}\sin(\lambda p^+x_c) f(\lambda).
\end{equation}
In the above equation the first term is the imaginary part of the S-matrix element,
the second
term is the real part and it can be estimated
with dispersion relations. Here we estimate it by another method.  We note
that the limit $s\to\infty$ implies $x_c\to 0$, the asymptotic behavior of
$g_h(x)$ with $x\to 0$ is expected to be
\begin{equation}
 xg_h(x) \sim x^{-\alpha},
\end{equation}
where $\alpha >0$. With this behavior one can expect that the dominant contribution
to the second term is determined by the asymptotic behavior.
It is clearly that this behavior is determined
by the behavior of $f(\lambda)$ for $\lambda\to\infty$. For
$\lambda\to\infty$ the function $f(\lambda)$ goes to zero, if $f(\lambda)$
converges to zero fast enough with $\lambda\to\infty$, the singular
behavior in Eq.(42) will not appear. If we calculate $f(\lambda)$ with perturbative
theory, the result looks like
\begin{equation}
f(\lambda)\vert_{\rm pert.} = \lambda^{-2}
\left( c_0 +\sum_{n=1} c_n (\ln (\lambda\mu))^n\right)
  + \cdots,
\end{equation}
where $\mu$ is the renormalization scale. The terms represented by
$\cdots$ will result in singular terms, like $\delta$-functions in the perturbatively calculated
gluon distribution. These terms are irrelevant in our case.
Neglecting the terms with $\ln(\lambda\mu)$,
$f(\lambda)$ behaves like the power behavior $\lambda^{-2}$. However, if one sums all
terms with the logarithm $\ln(\lambda\mu)$, the power behavior will be changed. Also
nonperturbative effects will definitely change the behavior. We assume that
the function $f(\lambda)$ takes the form for $\lambda\to\infty$:
\begin{equation}
f(\lambda)\approx f_{as}(\lambda) = \frac{a_g}{\lambda^\beta}
\end{equation}
and rewrite the integral in the definition of $g_h(x)$ as:
\begin{equation}
 xg_h(x) =-\frac{2}{p^+}\int^\infty_0
\frac{d\lambda}{2\pi} \cos(\lambda p^+ x)f_{as}(\lambda)
-\frac{2}{p^+}\int^\infty_0
\frac{d\lambda}{2\pi} \cos(\lambda p^+ x)(f(\lambda)-f_{as}(\lambda)).
\end{equation}
In the second term the function $(f(\lambda)-f_{as}(\lambda))$ will converge to
zero faster than $f_{as}(\lambda)$, when $\lambda\to \infty$,
hence the most singular term of $xg_h(x)$
for $x\to 0$ comes from the first term. We write the distribution as:
\begin{equation}
xg_h(x) =G_{as}(x)\cdot\left( 1+G_r(x)\right ), \ \ \  G_{as}(x)=\frac{A_g}{x^\alpha},
\end{equation}
where $G_r(x)\to 0$ with $x\to 0$. It should be noted that in general
$G_r(x)\cdot G_{as}(x)$ can be singular for $x\to 0$,
$G_{as}(x)$ is the most singular term in $g_h(x)$ for $x\to 0$.
With these notations the most singular term
in $g_h(x)$ is related to $f_{as}(\lambda)$:
\begin{eqnarray}
G_{as}(x) &=& -\frac{2}{p^+}\int^\infty_0
\frac{d\lambda}{2\pi} \cos(\lambda p^+ x)f_{as}(\lambda)
\nonumber\\
  &=& -\frac{2}{p^+} (xp^+)^{-1+\beta}a_g\int^\infty_0
   \frac{d\lambda}{2\pi} \lambda^{-\beta} \cos (\lambda).
\end{eqnarray}
It should be noted that the integral is finite as discussed in the appendix.
Comparing with $G_{as}(x)$ in Eq.(47) we obtain the relation between $\alpha$ and $\beta$
and that between $A_g$ and $a_g$:
\begin{eqnarray}
 \alpha &=& 1 - \beta, \nonumber\\
 A_g &=&  -2 (p^+)^{-2+\beta}a_g\int^\infty_0
   \frac{d\lambda}{2\pi} \lambda^{-\beta} \cos (\lambda).
\end{eqnarray}
\par
With the above discussion we can realize that the dominant contribution to $T_R $
is obtained by replacing $f(\lambda)$ with $f_{as}(\lambda)$, and this
dominant contribution is determined by the most singular term in the
gluon distribution. We will only take this dominant contribution. With
the form of $f_{as}(\lambda)$ it is straightforward to calculate $T_R $
for $t\to 0$:
\begin{eqnarray}
\FT
  &\approx &  (2\pi i)g_s^2 \frac{k^-}{4m_c}\int_0^\infty \frac{d\lambda}{2\pi}
 e^{-i \lambda x_c p^+} f_{as}(\lambda)
\nonumber\\
 & =&  i g_s^2 \frac{k^-}{4m_c}(x_c p^+)^{-1+\beta} a_g \int_0^\infty {d\lambda}
  \lambda^{-\beta} \left ( \cos(\lambda) -i \sin (\lambda)\right).
\end{eqnarray}
With the relations in Eq.(48) and results in the appendix for the integrals we
obtain:
\begin{equation}
\FT \approx -i\pi g_s^2 \frac{s}{8m_c} G_{as}(x_c) \left[ 1
  - i \tan(\frac{1}{2}\alpha\pi) \right].
\end{equation}
For a enough small $x_c$ one may replace $G_{as}(x_c)$ with
$[x_cg_h(x_c)]$ as a good approximation:
\begin{equation}
\FT \approx -\frac{i\pi}{4} g_s^2 m_c g_h(x_c) \left[ 1
  - i \tan(\frac{1}{2}\alpha\pi) \right].
\end{equation}
In the above result the spin of the light hadrons are the same.
With the S-matrix element the forward differential cross section reads:
\begin{eqnarray}
 \frac{d\sigma}{dt}\vert_{t\to 0} & = & \frac{2}{3\pi s^2}
 \frac{\Gamma (J/\psi\to e^+e^- )}{\alpha_{em}} \frac{1}{m_c^3}
  \overline{\sum} \vert T_R \vert ^2
 \nonumber\\
  &\approx & \frac{2 \pi^3\alpha_s^2}{3}\frac{\Gamma (J/\psi\to e^+e^- )}
  {\alpha_{em} m_c s^2} \left\vert g_h(x_c)(1- i \tan(\frac{1}{2}\alpha\pi))\right\vert^2
\end{eqnarray}
where $\overline{\sum}$ is the summation over spin of the final hadron $h$ and
the spin average of the initial hadron.
This result is our main result in this section. It should be emphasized that
we only used the asymptotic behavior of $g_h(x)$ to estimate the real part
of the amplitude, i.e., the term related to $\tan(\frac{1}{2}\alpha\pi)$, and the imaginary part is determined without using the asymptotic behavior,
as it already stands in Eq.(41). With the result in Eq.(51) and Eq.(52) a problem
may arise if $\alpha\to 1$. If $\alpha$ is really close to $1$ or equals to $1$, then
the cross section will become infinitely large. However, this can not be the case
because $\alpha =1$ implies that the second moment of $g_h(x)$, which is the average
of the momentum fraction carried by a gluon in the hadron $h$, is infinitely large.
Experimentally the extracted second moment is smaller than $1$. The value of $\alpha$
from different parameterization of $g_h$ ,
relevant to our case, is found to be $0.15 \sim 0.35$ for $\jpsi$,
the corresponding value for $\Upsilon$ production
is $0.30 \sim 0.48$.
\par
Now we are in the position to discuss the assumption leading to the
approximation in Eq.(37) and the mentioned ambiguity. After the integration over
$z$ in Eq.(34) we have neglected the $i0^+$. Keeping this term
the result reads:
\begin{equation}
T_R
 \approx (2\pi i)g_s^2 \frac{k^-}{4m_c}\int_0^\infty \frac{d\lambda}{2\pi}
 e^{-i \lambda x_c p^+-0^+\lambda} \langle h(p+\Delta )\vert G^{a,+\rho}(\frac{\lambda}{2} n)
  G^{a,+}_{\ \ \ \rho} (-\frac{\lambda}{2} n) \vert h(p)\rangle.
\end{equation}
With this term we can expand the matrix element in $\lambda$ and
perform the integration over $\lambda$ analytically. The result is:
\begin{equation}
T_R
 \approx \pi g_s^2 \frac{s}{4m_c} \sum_{n=0}^{\infty}
    \frac {1}{(x_c-i0^+)^{2n+1}}\frac{
    \langle h(p+\Delta)\vert O_{2n+2} \vert h(p) \rangle}
    {(p^+)^{2n+2}},
\end{equation}
where operators $O_J$ with $J=2,4,6,\cdots$ are the standard twist-2 operators in
the light-cone gauge:
\begin{equation}
 O_J = G^{a,+\mu}\left( \frac{i}{2}\hflrpt
 \right)^{J-2}G^{a,+}_{\ \ \ \mu},\ \ \ \ J=2,4,6,\cdots,
\end{equation}
and the $J$-th moment $a_J$ of gluon distribution function is then given by:
\begin{equation}
 \langle h(p)\vert O_J \vert h(p) \rangle =(p^+)^J a_J.
\end{equation}
In the following discussion we
neglect the spin-flip terms, these terms can be discussed in the same way
and they do not contribute at the leading order.
\par
In the limit $s\gg 4m_c^2\gg t$ we can first neglect the $t$-dependence
of matrix elements and only keep the dependence of $p^+$ and of $\Delta^+$.
For $\Delta^+\to 0$ the matrix elements can be expanded as:
\begin{equation}
\frac{
    \langle h(p+\Delta)\vert O_{2n+2} \vert h(p) \rangle}
    {(p^+)^{2n+2}} = a_{2n+2} +\sum_{k=1}^{\infty}
     (\frac{-\Delta^+}{2p^+})^k b_{2n+2,k}
      =a_{2n+2} +\sum_{k=1}^{\infty}x_c^k b_{2n+2,k},
\end{equation}
using these expansions we obtain
\begin{equation}
T_R
  \approx \pi g_s^2 \frac{s}{4m_c}\left\{
  \sum_{n=0}^{\infty} \frac {1}{(x_c-i0^+)^{2n+1}} a_{2n+2}
  +  \sum_{k=1}^{\infty} x_c^k \sum_{n=0}^{\infty}
   \frac {1}{(x_c-i0^+)^{2n+1}} b_{2n+2,k}\right\}.
\end{equation}
If we assume that for $n\to \infty$ the behavior of $b_{2n+2,k}$
is similar as $a_{2n+2}$ or $b_{2n+2,k}$ converges faster than
$a_{2n+2}$, then the dominant contribution for small $x_c$ comes
from the first sum, i.e., the dominant contribution comes from
the terms with the moments $a_J$ of the gluon distribution,
and other sums with $b_{2n+2,k}$ are suppressed by positive
powers
of $x_c$. The first sum can be written as a integration from, which
is just the approximation used in Eq.(37).
Now we consider the approximation in Eq.(39). Using translational
covariance the integral can be written:
\begin{equation}
T_R
 \approx (2\pi i)g_s^2 \frac{k^-}{4m_c}\int_0^\infty \frac{d\lambda}{2\pi}
 e^{-i 2\lambda x_c p^+-0^+\lambda} \langle h(p+\Delta )\vert G^{a,+\rho}(0)
  G^{a,+}_{\ \ \rho} (-\lambda n) \vert h(p)\rangle.
\end{equation}
Similarly, the integral can be written as a sum:
\begin{equation}
T_R
 \approx \pi g_s^2 \frac{s}{4m_c} \sum_{n=0}^{\infty}
    \frac {1}{(2x_c-i0^+)^{n+1}}\frac{
    \langle h(p+\Delta)\vert G^{a,+\mu}(i\partial^+)^n G^{a,+}_{\ \ \ \mu} \vert h(p) \rangle}
    {(p^+)^{n+2}}.
\end{equation}
We divide the sum into a parts with even $n$ and another part with odd $n$:
\begin{eqnarray}
T_R
 &\approx& \pi g_s^2 \frac{s}{4m_c}  \sum_{k=0}^{\infty}
    \frac {1}{(2x_c-i0^+)^{2k+1}}\frac{
    \langle h(p+\Delta)\vert G^{a,+\mu}(i\partial^+)^{2k} G^{a,+}_{\ \ \ \mu} \vert h(p) \rangle}
    {(p^+)^{2k+2}}
    \nonumber\\
    && + \pi g_s^2 \frac{s}{4m_c}    \sum_{k=0}^{\infty}
    \frac {1}{(2x_c-i0^+)^{2k+2}}\frac{
    \langle h(p+\Delta)\vert G^{a,+\mu}(i\partial^+)^{2k+1} G^{a,+}_{\ \ \ \mu} \vert h(p) \rangle}
    {(p^+)^{2k+3}}.
\end{eqnarray}
For $\Delta^+\to 0$ the matrix elements behave like:
\begin{eqnarray}
\frac{
    \langle h(p+\Delta)\vert G^{a,+\mu}(i\partial^+)^{2k} G^{a,+}_{\ \ \ \mu} \vert h(p) \rangle}
    {(p^+)^{2k+2}} &=& a_{2k+2} +{\cal O} (\frac{\Delta^+}{p^+}),
    \nonumber\\
  \frac{ \langle h(p+\Delta)\vert G^{a,+\mu}(i\partial^+)^{2k+1} G^{a,+}_{\ \ \ \mu} \vert h(p) \rangle}
    {(p^+)^{2k+3}} &=& -(2k+1)\frac{\Delta^+}{2p^+} a_{2k+2} +{\cal O} ((\frac{\Delta^+}{p^+})^2)
    \nonumber\\
                   &=& (2k+1)x_c a_{2k+2} +{\cal O} (x_c^2)
\end{eqnarray}
From these one would neglect the matrix elements with odd numbers of derivatives,
i.e., one would neglect the summation in the
second line of Eq.(61),  because the matrix elements with odd numbers of derivatives
go to zero with $\Delta^+\to 0$, or with $x_c\to 0$ by noting $\Delta^+=-2x_cp^+$.
Keeping only the leading terms of the matrix elements with even numbers
of derivatives and neglecting the matrix elements with odd numbers of derivatives,
one obtains:
\begin{eqnarray}
T_R \approx \pi g_s^2 \frac{s}{4m_c}  \sum_{k=0}^{\infty}
    \frac {1}{(2x_c-i0^+)^{2k+1}} a_{2k+1}.\nonumber
\end{eqnarray}
Writing the sum into a integration form as done for Eq.(58),
one obtains the approximation in Eq.(39) and
we will get the result for $T_R$ which is
related to $g_h(x)$ with $x=2x_c=4m_c^2/s$.
But, the matrix elements with odd numbers of derivatives in the second
line in Eq.(61) can not be neglected. Although they are proportional to $x_c$ as $x_c\to 0$, but
the power of $x_c$ in the denominator in the second sum is $2k+2$ in comparison with
the power $2k+1$ in the sum of the first line. Keeping the leading terms
for matrix elements with even numbers and odd numbers of derivatives, one obtains
\begin{eqnarray}
T_R\approx  \pi g_s^2 \frac{s}{4m_c}  \sum_{k=0}^{\infty}
    \frac {1}{(2x_c-i0^+)^{2k+1}}\cdot a_{2k+1} +
 \pi g_s^2 \frac{s}{8m_c}  \sum_{k=0}^{\infty}\frac{(2k+1)}
    {(2x_c-i0^+)^{2k+1}} a_{2k+1}. \nonumber
\end{eqnarray}
Clearly both sums give equally important
contributions to $T_R$. Actually every term in the expansion in $\Delta^+$
for matrix elements with odd numbers of derivatives can lead to a equally
important contribution to $T_R$.
To sum the leading order
contributions one can use
\begin{equation}
\langle h(p+\Delta)\vert G^{a,+\mu}(\partial^+)^n G^{a,+}_{\ \ \ \mu} \vert h(p) \rangle
= \sum_{k=0}^{n} \frac{n!}{k!(n-k)!}
  (\frac{i\Delta^+}{2})^k
\langle h(p+\Delta)\vert G^{a,+\mu}
\left(\frac{1}{2}\hflrpt\right)^{n-k} G^{a,+}_{\ \ \ \mu} \vert h(p) \rangle
\end{equation}
to re-arrange the sum. With the same assumption made after Eq.(58) one will get
the same result as that obtained with the approximation in Eq.(37).
\par
In this section we neglect corrections suppressed by the inverse of $s$ in the limit
$s\to\infty$ and
use two assumptions to derive our results in Eq.(51) and Eq.(52): One is used in Eq.(37)
and specified in detail between Eq.(58) and Eq.(59), another is to use the assumed
asymptotic behavior to determine the real part of the S-matrix element.
Without these two assumptions  one can relate
$F_R(z)$ in the limit to the skewed gluon distribution
by neglecting corrections suppressed by the inverse of $s$. The relation
reads:
\begin{equation}
F_R(z)\approx -g_s^2 \frac{s}{8m_c} x_1x_2 f_g(x_1,x_2),
\end{equation}
where we used the definition of skewed parton distributions in \cite{JC}.
There exists different definitions and the relation between them is well
discussed in \cite{Ji2}.
The variables $x_1$ and $x_2$ are related to $z$ as:
\begin{equation}
x_1=\frac{2m_c^2}{s}(1-z),\ \ \ \ \ \ x_2=-\frac{2m_c^2}{s}(1+z).
\end{equation}
With this we end up with the result for $T_R$:
\begin{equation}
T_R \approx  -\frac{g_s^2m_c}{4} \int dx_2 \frac{1}{(x_1-i0^+)(x_2+i0^+)}
 x_1x_2 f_g(x_1,x_2),
\end{equation}
with
\begin{equation}
x_1-x_2 =\frac{4m_c^2}{s}.
\end{equation}
Without any knowledge about $f_g(x_1,x_2)$, the integral can not be performed and
numerical predictions can not be made.
\par
With our results  we are in position
to compare our result with others obtained in \cite{FKS}. The starting point
there is to consider the process in Eq.(1), where the initial photon is
virtual. The interaction between the initial hadron
$h$ and the $c\bar c$ pair is taken as that between $h$ and a $c\bar c$ dipole
with a small transverse-size. Using the perturbative result, in which
only the exchange of two gluons are considered, one obtains at the leading
order of $Q^{-2}$:
\begin{equation}
\frac{d\sigma}{dt}\vert_{t\to 0} \sim \frac{1}{(Q^2)^4} \vert
         (1+i\beta_s ) \left [ x g_h(x)\right ]\vert^2 \left\{
          1+{\cal O}(Q^{-2})\right\},  \ \ \ \ {\rm with}
         \ x=\frac{Q^2}{s}.
\end{equation}
To derive the result one uses a light-cone wave function to
describe the nonperturbative properties of $J/\psi$ and $J/\psi$
should be taken as massless to consistently perform the expansion
in $Q^{-2}$. In the above equation the terms relevant for our
comparison are given explicitly. As also pointed in \cite{SJB},
the leading log approximations in $\ln (1/x)$ and in $\ln
(Q^2/\Lambda^2_{QCD})$ are required to identify the quantity
$x=Q^2/s$ as the variable of the gluon distribution $g_h(x)$. The
produced $J/\psi$ is longitudinally polarized. It should be noted
that by taking the $c\bar c$ pair as a small dipole it is
equivalent to neglect higher orders in $Q^{-2}$\cite{FKS}. To
include the production of a transversally polarized $J/\psi$, the
above result is re-derived by keeping the mass $M_{J/\psi}$ and it
is found that at leading order of $Q^{-2}$ the result is obtained
by replacing $Q^2$ with $Q^2+M^2_{J/\psi}$. The results for the
production of $J/\psi$ with a real photon are obtained simply by
taking $Q^2\to 0$ in the above result, after the replacement of
$Q^2$ with $Q^2+M^2_{J/\psi}$\cite{FKS,FDS}. Several modifications
of the results are then introduced, and the relation between the
forward S-matrix element and the gluon distribution becomes
complicated. We will compare the result in \cite{MR,FKS,FDS}
without these modifications. The result in \cite{FKS,FDS} by
taking $Q^2\to 0$ is:
\begin{equation}
\frac{d\sigma}{dt}\vert_{t\to 0} \sim \frac{1}{(4m^2_c)^4} \vert
         (1+i\beta_s ) \left [ x g_h(x)\right ]\vert^2,  \ \ \ \ {\rm with}
         \ x=\frac{4m_c^2}{s},
\end{equation}
where the term with $\beta_s$ is the real part of the amplitude and is given by:
\begin{equation}
 i\beta_s \left [ x g_h(x)\right ] \approx i\frac{\pi}{2}
     \frac {\partial xg_h(x)}{\partial \ln x }.
\end{equation}
If we take $\alpha$ in Eq.(42) as a small parameter, this part is the same as
ours.
The main difference between our result and that given above is that the
S-matrix element is related to the gluon distribution $g_h(x)$ at
different $x$. It is possible that the limit $Q^2\to 0$ can not
be taken because higher orders in $Q^{-2}$ are neglected in Eq.(68).
\par
Similar derivation of the results by using a nonrelativistic wave-function for $J/\psi$
is also done in \cite{MR,MRT}, by keeping the mass of $J/\psi$ at the leading order
of $Q^{-2}$ and neglecting higher orders in $Q^{-2}$. Setting $Q^2=0$ one obtains:
\begin{equation}
 \frac{d\sigma}{dt}\vert_{t\to 0}
  \approx  \frac{16\pi^3\alpha_s^2}{3}\frac{\Gamma (J/\psi\to e^+e^- )}
  {\alpha_{em} M^5_{J/\psi} }\left ( 2x_c g_h(2x_c)\right )^2.
\end{equation}
This result looks similar as ours, but the differential cross section is related
to $x g_h(x)$ at $x=2x_c$ as the result in Eq.(69) and the real part of the amplitude
is neglected.

\par
\vskip20pt
\noindent
{\bf 4. Comparison with experiment}
\par\vskip20pt
In this section we will compare our results with experiment performed
at HERA. In the last two sections we have given our results, but not
specified the energy scale $\mu$, at which the nonperturbative quantities
like the gluon distribution are defined. To compare with experiment
one must choose this scale. In general one may take the scale $\mu$
as a soft scale at order of $\Lambda_{QCD}$ because of the emission of
soft gluons. If one takes
$\mu$ as a soft scale, then the scale should be the same in the production
of $\Upsilon$. However, there can be exchanges of hard gluons
between quark lines, between gluon lines and between quark- and gluon lines.
The effect can be studied with perturbative QCD and log terms like $\ln (\mu/m_c)$
appear, one may then identify the scale
$\mu$ as $m_c$ for $J/\psi$ and as $m_b$ for $\Upsilon$, respectively.
We will compare our results with experiment with these two choices.
It should be pointed out that significant corrections to our results for
$J/\psi$ exist, as discussed below. Before these corrections are under
control, a detailed comparison of theoretical results with experiment
can not be made. In this section we only give numerical results
based on our results at leading orders without any modifications, suggested
by those corrections.
\par
\begin{figure}[htb]
\centering
\includegraphics[width=10cm,height=8cm]{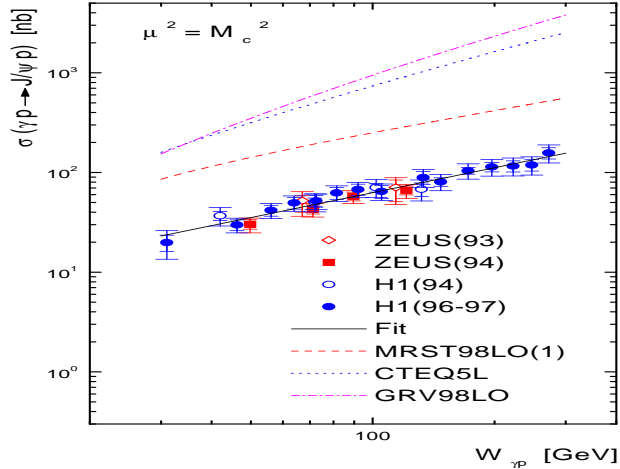}
\vspace*{-16mm}
\caption{The cross section of elastic photoproduction of $\jpsi$ versus $W_{\gamma p}$,
the photon-proton center-of-mass energy. The data points are published HERA
results\cite{Herajpsi}. The full solid line represents a fit of the
form $\sigma (\jpsi) \propto W_{\gamma p}^{\delta}$.
The theoretical predictions of present work using various parameterizations of
the leading gluon density in the proton at $m_c$ scale are also shown.}
\label{jpsidata}
\end{figure}

To predict the total cross section we assume that
\begin{equation}
\frac{d\sigma}{dt}(t) = \frac{d\sigma}{dt}(t=0) e^{-B\vert t\vert},
\end{equation}
where the slope parameter $B$ is measured as
$B\approx 4.5{\rm GeV}^{-2}$\cite{Herajpsi}.
If the scale $\mu$ is taken as a soft scale, there is no detailed information
available for the gluon distribution at the scale, one may know that
$xg_h(x)$ is divergent like $x^{-\alpha}$ when $x\to 0$. Then the cross
section behaves like:
\begin{equation}
\sigma(J/\psi)= A \left ( \frac{W_{\gamma p}}{W_0}\right )^\delta,\ \ \ \ W_0=1{\rm GeV},
\end{equation}
where $W_{\gamma p}=\sqrt{s}$ and $\delta=4\alpha$.
If we assume that the parameter $B$ is the same for $\Upsilon$, then we can
predict the cross section of $\Upsilon$ by determining the parameters in
Eq.(73) from experimental data for $J/\psi$\cite{Herajpsi}. This can also be
considered as we neglect the $\mu$-dependence.
We fit the published HERA data with
Eq.(73) and obtain:
\begin{equation}
\delta =0.83\pm 0.13, \ \ \ \ \, A=1.37\pm 0.84{\rm nb},
\ \ \ \ \ \chi^2/d.o.f. =0.31.
\end{equation}
The fitting quality is quite good indicated by the small value of
$\chi^2/d.o.f.$. The fitting cure with experimental data is also
shown in Fig. \ref{jpsidata}. With the determined parameters we can predict:
\begin{eqnarray}
\sigma(\Upsilon) &\approx & 2.7 \times 10^{-2}~{\rm nb},
\ \ \ \, {\rm for}\ W_{\gamma p}=120 ~{\rm GeV},
\nonumber\\
\sigma(\Upsilon) &\approx & 3.2 \times 10^{-2}~{\rm nb},
\ \ \ \, {\rm for}\ W_{\gamma p}=143 ~{\rm GeV}.
\end{eqnarray}
However these predicted values are too small in comparison with the central values
of experimental results shown in Fig. \ref{updata}. Different reasons are responsible for this
discrepancy. Firstly we note that there are large errors in experimental data,
even large errors in determining the energy $W_{\gamma p}$. Secondly, there are several
uncertainties in our predictions, e.g., relativistic corrections arising
in the expansion in Eq.(12), corrections from higher orders in the
expansion in $m_c^{-1}$, and also the leading order determination of the
matrix element through the leptonic decay in Eq.(17), etc..
Among them the effects from relativistic corrections and from the determination
of the matrix element can be most significant. Because the $c$- or $\bar c$ quark
moves in the $J/\psi$ rest frame with the velocity $v_c$, which is not very small,
indicated by $v_c^2 \approx 0.3$, the relativistic correction can be significant.
In our approach this correction can be systematically added and its study
is underway. In the determination of the matrix element we have used
the leading order result for the leptonic decay in Eq.(17). It is well
known that the corrections from higher orders in $\alpha_s$ are large\cite{ML,BSS}.
Also, the relativistic correction to the result is significant\cite{Rma}.
It is clear that without these corrections an accurate prediction can not be made
for the production of $J/\psi$. For the production of $\Upsilon$ these corrections
are expected to be small.

\begin{figure}[htb]
\centering
\includegraphics[width=10cm,height=8cm]{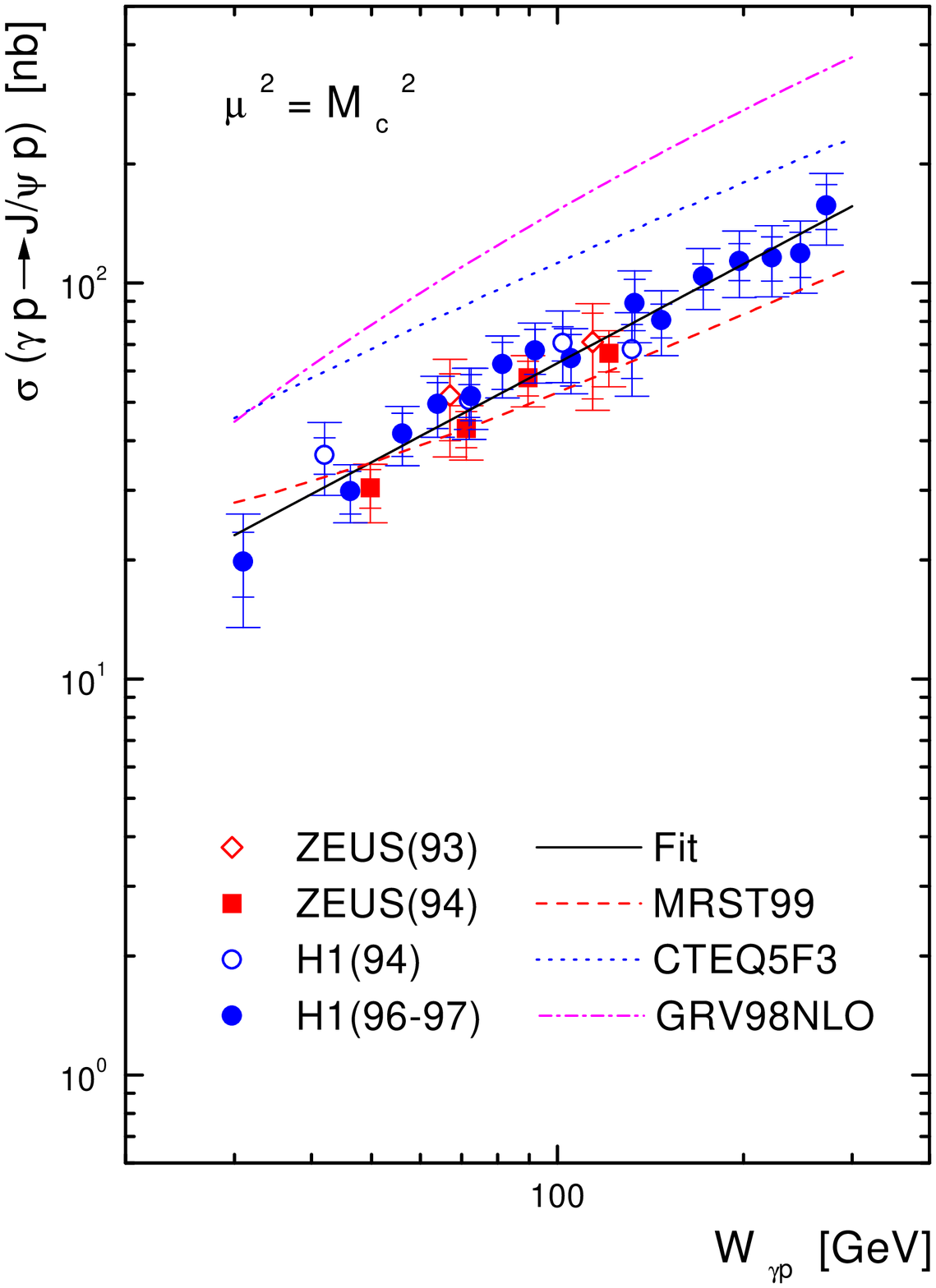}
\vspace*{-16mm}
\caption{The cross section of elastic photoproduction of $\jpsi$ versus $W_{\gamma p}$,
the photon-proton center-of-mass energy. The data points are published HERA
results\cite{Herajpsi}. The full solid line represents a fit of the
form $\sigma (\jpsi) \propto W_{\gamma p}^{\delta}$.
The theoretical predictions of present work using various parameterizations of
the next-to-leading gluon density in the proton at $m_c$ scale are also shown.}
\label{jpsinlo}
\end{figure}

\par
If we take the scale $\mu$ as the heavy quark mass, we can predict cross sections
without any input parameter, because the gluon distribution at the scale is
determined. We use the recently determined sets of parton distributions GRV98\cite{GRV},
CTEQ5\cite{CTEQ}, MRST99\cite{MRST99} and MRST98LO\cite{MRST98LO}.
We do not use the newest MRST01 distributions for $J/\psi$, because
it has an unphysical zero at $\mu=m_c$ in the gluon distribution
and for small $x$ the distribution turns
to be negative\cite{MRST}. Instead we use an old set of parton distributions\cite{MRST99}.
The predictions with experimental data\cite{Herajpsi} are also shown in Fig.2 with leading order
gluon distribution, and  in Fig.3 with the next-to-leading order gluon distribution.
With these distributions we can determine the contribution of the real part of the S-matrix element
to the cross section. The contribution is at $12\%$-, $20\%$- and $26\%$ level with the leading order gluon
distribution of MRST98LO, CTEQ5L and GRV98LO, respectively, while with the next-to-leading order
gluon distribution of MRST99, CTEQF3 and GRV98 it is at $6\%$-, $8\%$- and $12\%$ level, respectively.
From these figures we can see that
all distributions give the results roughly with the same feature that the cross section
increases with the energy, numerical results with the next-to-leading distribution of MRST99
are close to the experimental results for large $s$,
while the prediction with other gluon distributions gives a rather
large cross section.
Although a rather good description of experimental data for $J/\psi$ can be given with the next-to-leading
order gluon distribution of MRST99,
one should keep in mind that our results can have significant corrections as discussed before and that
it is not consistent to use next-to-leading order gluon distributions with our results at leading order
of $\alpha_s$. We note that the cross section
does not depend on the renormalization scale $\mu$. Since we work only at $\alpha_s^2$-the leading
order, the $\mu$-dependence can appear at the next-to-leading order of
$\alpha_s$, i.e., at $\alpha_s^3$. To use next-to-leading order gluon distributions, the corrections from the next-to-leading
order of $\alpha_s$ should be also included and the $\mu$-dependence at $\alpha_s^3$ is eliminated. These corrections come from exchanges of hard gluons
between quarks and gluons in Fig.1. and they can be large because of that $\alpha_s$ at $\mu =m_c$ is not
a very small parameter. Unfortunately, these corrections are unknown. It is clear that a conclusive comparison
without those corrections, i.e., the relativistic correction and one-loop correction, can not be made with
experiment.
\par
\begin{figure}[htb]
\centering
\includegraphics[width=10cm,height=8cm]{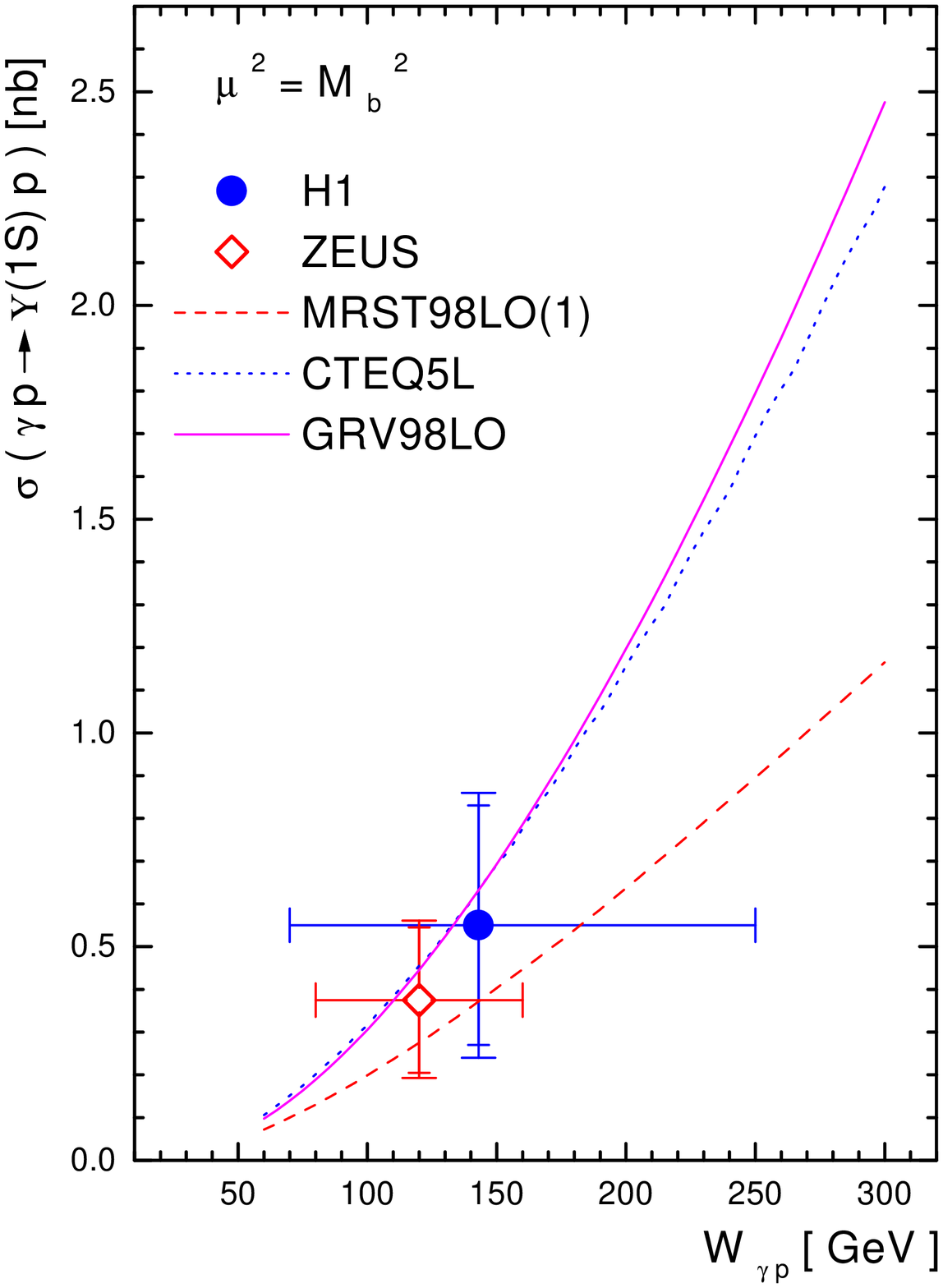}
\vspace*{-16mm}
\caption{The cross section of elastic photoproduction of $\Upsilon (1S)$ versus
$W_{\gamma p}$, the photon-proton center-of-mass energy.
The data points are published HERA
results\cite{Heraup}.
The theoretical predictions of present work using various parameterizations of
the gluon density in the proton at $m_b$ scale are also shown.}
\label{updata}
\end{figure}

\begin{figure}[htb]
\centering
\includegraphics[width=10cm,height=8cm]{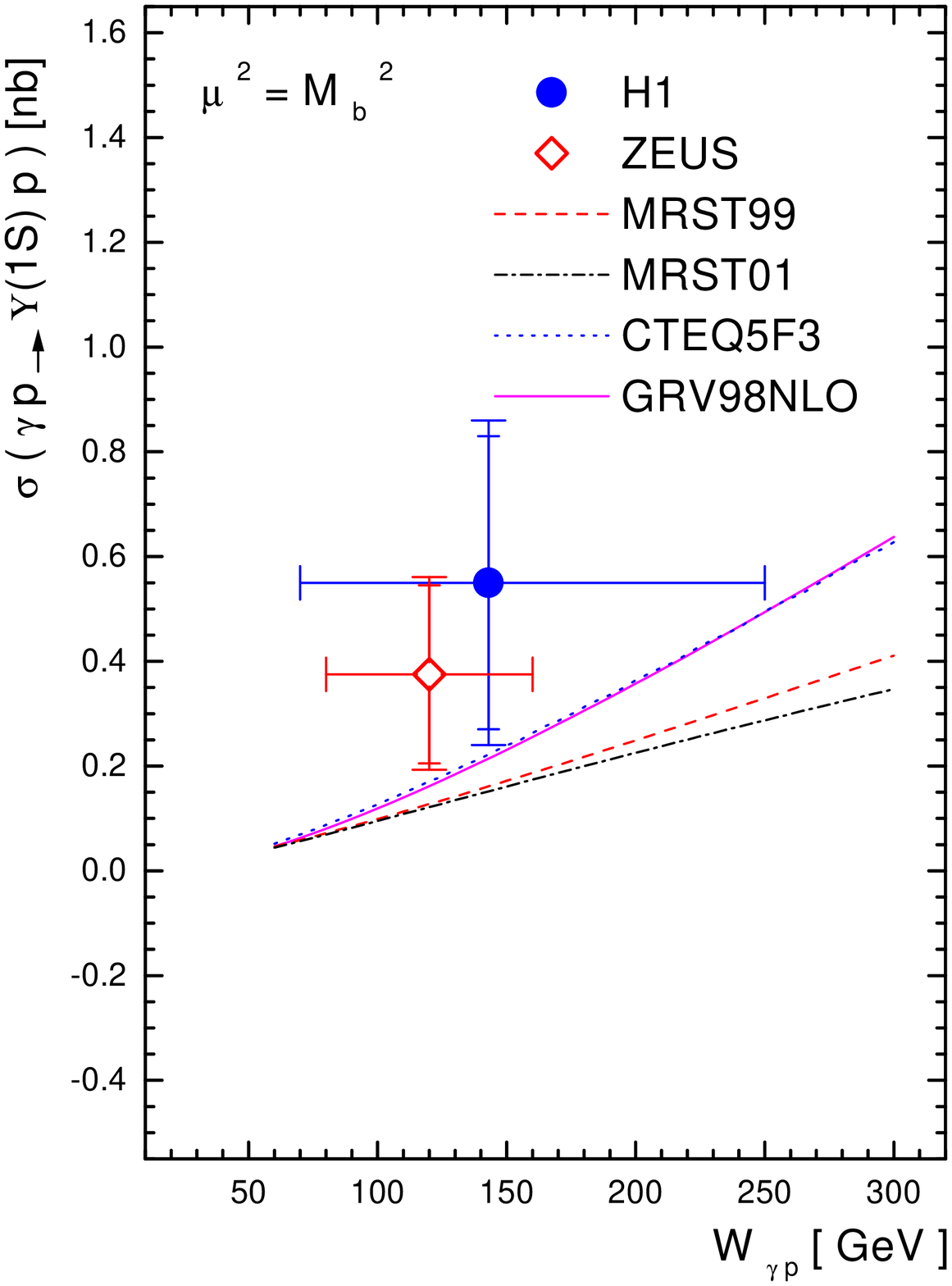}
\vspace*{-16mm}
\caption{The cross section of elastic photoproduction of $\Upsilon (1S)$ versus
$W_{\gamma p}$, the photon-proton center-of-mass energy.
The data points are published HERA
results\cite{Heraup}.
The theoretical predictions of present work using various parameterizations of
the gluon density in the proton at $m_b$ scale are also shown.}
\label{upnlo}
\end{figure}
For $\Upsilon$ our predictions are obtained by taking $\mu =m_b =5$GeV. The
discussed corrections should be small for the case with $\Upsilon$. We indeed find
that our prediction with leading order gluon distributions is in agreement
with experiment, although the experimental errors are large.
Our numerical results are shown
with experimental data\cite{Heraup}
in Fig.4 with leading order
gluon distributions, and  in Fig.5 with the next-to-leading order gluon distributions.
The contribution of the real part of the amplitude to the cross section
is at $36\%$-, $43\%$- and $47\%$ level with the leading order gluon distribution
of MRST98LO, CTEQ5L and GRV98LO, respectively. With the next-to-leading order
gluon distribution of MRST01, MRST99, CTEQF3 and GRV98 it is at $20\%$-,
$23\%$-, $29\%$- and $33\%$ level, respectively.
For $\Upsilon$ productions another possible correction besides those discussed before
is that at $\mu =m_b$ the finite skewness, which we have neglected to relate
the S-matrix element to the gluon distribution, can lead a significant effect. Studies in \cite{MR2,UFS}
show that this is the case. Instead using gluon distribution one may need to use
the skewed gluon distribution to make numerical predictions. However, this distribution
is not well known and numerical predictions cannot be made. It deserves
a further study of corrections to our results both for $J/\psi$ and for $\Upsilon$.

\par
\vskip20pt
\noindent
{\bf 5. Summary}
\par\vskip20pt
In this work we have studied diffractive photoproduction of $J/\psi$, where we have taken
the gluons exchanged between the initial hadron and the $c\bar c$ pair as soft gluons.
We then have used an expansion in $m_c^{-1}$ to study the effect of the exchange of soft
gluons. Our results have been derived with the assumption the exchange of two gluons in a special gauge.
But, they can also be derived without the assumption in an arbitrary
gauge. This can be done by formulating the problem
as the splitting of the photon into a $c\bar c$ pair in a background field
of gluons, which varies slowly in the space-time.
Our results for the S-matrix element consist of a NRQCD matrix element representing
the nonperturbative effect related to $J/\psi$ and a matrix element of two gluonic field strength
operators which are separated in the moving direction of $J/\psi$ in the space-time.
The matrix element of the field strength operators characterize the nonperturbative effect
related to the initial hadron. In the limit of $s\to\infty$ the forward S-matrix element
is related to the skewed gluon distribution. With a reasonable assumption the forward
S-matrix element can also be related to the usual gluon distribution, this enables us
to make numerical predictions and to compare with experiment.
\par
Since we take the exchanged gluons as soft gluons, it is not very clear how to identify the
renormalization scale $\mu$ to make numerical predictions. However, with the knowledge about
the asymptotic behavior of the gluon distribution $g_h(x)$ with $x\to 0$ our
results show that the total cross section increases with increasing energy. We first take
$\mu$ as a soft scale and fit experimental results for $J/\psi$ with our results.
The experimental data show clearly that the cross-section increases with the energy as
$W^\delta$. With determined parameters we can predict the cross section for $\Upsilon$.
However, the predicted cross sections are too small. Possible reasons are large corrections
to the leading order results for $J/\psi$. We also take $\mu$ as a hard scale, i.e.,
as the mass of the heavy quark, with an existing gluon distribution we are able to give
a rather good description of describe
experimental data for $J/\psi$, although large corrections are neglected. These
corrections for $\Upsilon$ are expected to be smaller those for $J/\psi$. We indeed
find that
our predictions with leading order gluon distributions for $\Upsilon$ production
are in agreement with experiment, but there are only two data points with large errors.
It deserves a further study of corrections to our results and of experiment.
\par
Our approach is distinctly different than previous approaches. In previous approaches
one start to use perturbative QCD to study diffractive production with a virtual
photon in the initial state, where the virtuality $Q^2$ of the photon is large
to ensure that perturbative QCD can be used for gluon exchange.
Keeping leading order in $Q^{-2}$ and
the finite mass of quarkonia one obtains results for the cross section. Then the cross
section of diffractive production of a real photon is obtained by setting $Q^2=0$.
Clearly, this setting can not be done properly, because higher orders in $Q^{-2}$
are neglected. In our approach we take the exchanged gluons as soft gluons and use
an expansion in the inverse of the heavy quark mass to handle the exchange of soft gluons.
Although the forward S-matrix element can be related to the gluon distribution as
that in the previous approaches, but the relation is different, as discussed in detail
in Sec. 3.
\par
In our approach corrections can be systematically added. Among them relativistic correction
can be most important for production of $J/\psi$, and a study of the relativistic correction
is underway for getting accurate results of theory. With the accurate results of theory
diffractive photoproduction may provide an interesting way to study the nonpertubative
nature of nucleon and of nuclei.

\vskip 5mm
\begin{center}
{\bf\large Acknowledgments}
\end{center}

The work of J. P. Ma is supported  by National Nature
Science Foundation of P. R. China and by the
Hundred Young Scientist Program of Academia Sinica of P. R. China,
the work of J. S. Xu is supported by the Postdoctoral Foundation of P. R. China and  by
the K. C. Wong Education Foundation, Hong Kong.


\vfil\eject
\par\vskip2cm
\noindent
{\bf Appendix}
\par\vskip20pt
In this appendix we study the integral which appears in Eq.(48). We define
\begin{equation}
I(\beta) =\int_0^\infty d\lambda \lambda^{-\beta} (\cos(\lambda) +i \sin(\lambda))
    =\int_0^\infty d\lambda \lambda^{-\beta} e^{i\lambda},
\end{equation}
where $\beta$ is real.
We make an exchange of variable $\lambda^{1-\beta} =z$, then the integral
becomes
\begin{equation}
I(\beta=1-\alpha) = \frac{1}{\alpha} \int_0^\infty dz \exp ( i z^{\frac{1}{\alpha}}).
\end{equation}

\begin{figure}[htb]
\centering
\includegraphics[width=5cm]{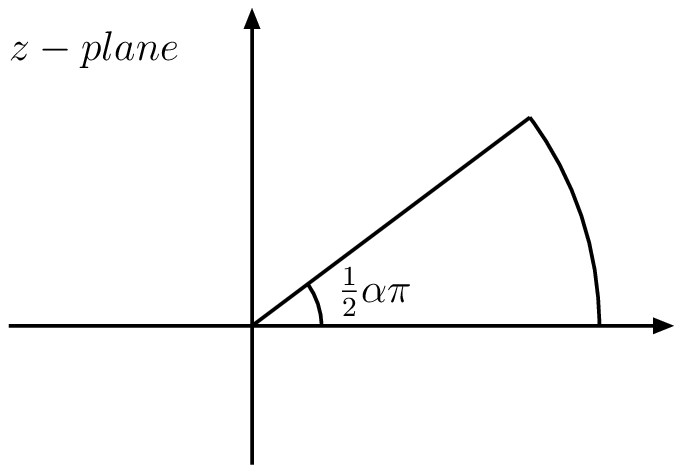}
\caption{The integration contour in complex $z-$plane.}
\label{zcont}
\end{figure}

\noindent
We extend the integrand as a complex function and consider
the contour integral where the contour in the complex $z$-plan
is specified as in Fig. \ref{zcont} with $\theta=\frac{1}{2}\alpha\pi$.
Using the contour integral we have:
\begin{equation}
I(\beta=1-\alpha) = -\frac{1}{\alpha} \exp(i\frac{1}{2}\alpha\pi)
 \int_0^\infty dR \exp(-R^{\frac{1}{\alpha}}),
\end{equation}
for $\alpha>0$ the integral is finite and it is real. With this result we obtain
\begin{equation}
\int_0^\infty d\lambda \lambda^{-\beta} \sin(\lambda)=\tan(\frac{1}{2}\alpha\pi)
 \int_0^\infty d\lambda \lambda^{-\beta} \cos(\lambda)
\end{equation}
which is used in Eq.(48).

\vfil\eject

\end{document}